\documentstyle[preprint,revtex]{aps}
\begin{document}
\hsize = 7.0in
\preprint{LA-UR-92-3133 and ISU-NP-92-11}
\widetext
\def\eq{\,=\,}
\def\beq{\begin{equation}}
\def\eeq{\end{equation}}
\def\l{\left(}
\def\r{\right)}
\def\po{\mathaccent 23 p}
\def\pp{p^+}
\def\pr{p_r}
\def\pl{p_\ell}
\draft
\begin{title}
Natural Hadronic Degrees of Freedom
\end{title}
\vskip -0.2in
\begin{title}
for an Effective
QCD Action in the Front Form \footnotemark\
\end{title} %
\footnotetext
{Part of this was done under the auspices of the U. S. Department of Energy.}
\author{~~~~~~~~~~~~~~~~~~~D. V. Ahluwalia \dag \ and Mikolaj Sawicki \ddag}
\begin{instit}
\dag Medium Energy Physics Division, MS H-850, Los Alamos
National Laboratory,\end{instit}
\vskip -0.1in
\begin{instit}
 Los Alamos, New Mexico 87544, USA.
\end{instit}
\begin{instit}
\ddag Department of Physics, Iowa State University, Ames, Iowa 50011, USA.
\end{instit}
\begin{abstract}
On the one hand any effective action of the QCD,
$S[\pi,\,\rho,\,\omega,\ldots,N,\,\overline N,\,\Delta^{3/2},\ldots]$,
would involve, at least in principle, a tower of hadronic fields
of all spins; on the other hand the front form of the field theory has
emerged as an appropriate framework to investigate the high energy
behaviour of many systems. Motivated by these simple considerations we
present a systematic formalism which explicitly provides the front
form fields appropriate for constructing a
$S[\pi,\,\rho,\,\omega,\ldots,N,\,\overline N,\,\Delta^{3/2},\ldots]$.
Further, a transformation matrix $\Omega(j)$ is constructed which
determines how the front form hadronic fields superimpose to yield the
more familiar instant form fields. Explicit results for the front form
hadronic fields up to spin two are catalogued in appendixes. For spin
one half, the $\Omega(1/2)$ coincides with the celebrated ``Melosh
transformation.'' The formalism, without explicitly invoking any wave
equations, reproduces spin one half front-form results of
Lepage and Brodsky, and Dziembowski.

\end{abstract}
\pacs{PACS numbers: 11.30.Cp, 11.80.Cr, 12.38.Bx}

\section{Introduction}
\label{sec:intro}

As presently understood, the fundamental constituents of the observed
hadrons are quarks and gluons. A great spectrum of nuclear phenomenon
are at energy/length scales where the empirically observed hadrons
appear to be the natural \cite{Walecka} degrees of freedom. As a new
generation of nuclear physics facilities become available at least two
aspects of nuclear matter will become more pronounced or observable.
First, the $j\,>\,{1\over 2}$ hadronic degrees of freedom will become
increasingly important. Second, a transition between hadronic matter
and the quark-gluon phase will offer new insights into nuclear matter
and the physical content of the QCD action. The first of the most
obvious question which one can ask is the relationship between these
two aspects. That is, does the transition {\it hadronic matter}
$\leftrightarrow$ {\it quark-gluon phase} exploit the high-spin
degrees of freedom provided by various hadronic resonances?  The
essential spirit of this point of view is contained in the work of
Cahill and collaborators \cite{Cahill} where the fundamental defining
action of QCD, $S[\overline q,\,q,\,A_\mu^a]$, is related via a change
of variables (using techniques of the functional integral calculus) to
an effective action for hadronic physics,
$S[\pi,\,\rho,\,\omega,\ldots,N,\,\overline N,\,\Delta^{3/2},\ldots]$.
This paper will provide, what in our opinion, are one of the most
natural objects for constructing
$S[\pi,\,\rho,\,\omega,\ldots,\overline N,\,N,\,\Delta^{3/2},\ldots]$.

We begin with the observation that since the simplest description
\cite{Walecka} of nucleons in nuclear matter requires the use of  Dirac
equation the nuclear phenomenon is {\it inherently} relativistic.  As
a result the front form of the quantum field theory is expected to
play an important role in nuclear and hadronic physics. The high-spin
hadronic degrees of freedom logically merge with this aspect --- for
{\it spin} emerges as a logical consequence of the Poincar\'e symmetry
\cite{Wigner}. In addition to the well known advantages of the front
form \cite{DiracLight,addLF,Soper,LeutwylerS} of the quantum field
theory we have recently gained some additional insights into the
nature of front-form field theory by studying the evolution of a quantum
system along an appropriately parameterised timelike \cite{da}
direction. By varying a single parameter the standard instant form and
front form results can be obtained \cite{ms} in a continuous and well
defined fashion for $j\,\le\,{1\over 2}$.  Because of the absence of a
unique wave equation and various problems which are encountered for
wave equations with $j\,>\,{1\over 2}$, these works do not easily
extend beyond spin one half. The clue to a possible solution to this
problem lies in a recent work where, following Weinberg
\cite{Weinberg} and Ryder \cite{Ryder}, one of the present authors in
collaboration with Ernst \cite{dva,prc,plb,mpl,wig} has explicitly
constructed instant-form Dirac-like spinors, Feynman-Dyson
propagators, and wave equations for arbitrary spins; and discovered
several interesting aspects of the Weinberg equations \cite{Weinberg}.
In contrast to the Rarita-Schwinger/Bargmann-Wigner
\cite{RaritaS,BargmannW} formalism, the instant-form work of Refs.
\cite{dva,prc} requires no constraints or auxiliary fields and in
addition incorporates the physically correct number of particle and
antiparticle spinorial degrees of freedom.

Here in this paper, following the logic of Refs.
\cite{Weinberg,Soper,dva,prc}, we
present a general procedure for constructing the
front form covariant spinors for
any spin {\it without} a direct reference to any wave equation. A new
transformation, $\Omega(j)$, which relates  the front form spinors to the
instant form spinors  will be constructed. Explicit examples
for $j\,=\,{1\over 2},\,1,\, {3\over 2}, \,2$ will be presented. For
$j\,=\,{1\over 2}$, the formalism reproduces the front form spinors of Lepage
and Brodsky \cite{lepage} and $\Omega({1\over 2})$ coincides with the
celebrated ``Melosh
transformation'' \cite{Melosh,Dziembowski}. Our work thus opens a possibility
to construct
$S[\pi,\,\rho,\,\omega,\ldots,N,\,\overline N,\,\Delta^{3/2},\ldots]$ in terms
of the front form fields obtained here, and provide it's connection with the
QCD action $S[\overline q,\,q,\,A_\mu^a]$.
This approach is in the
spirit of Dirac's \cite{DiracSpin} original motivation for the high-spin wave
equations for ``approximate application to composite particles'' and provides a
very natural framework for constructing a truly QCD based effective field
theory of hadronic resonances along the lines of recent work of Cahill {\it et
al.} cited above. Here we combine Dirac's original motivation just mentioned
with another of his classic ideas of constructing the front form of field
theory \cite{DiracLight} for high energy physics.
The present work also provides the unifying link
between Weinberg's high spin work \cite{Weinberg} and his well
known \cite{WeinbergI} paper on the infinite momentum frame.

Unless otherwise indicated we follow the notation of Refs. \cite{Ryder,prc}.
We use the notation $x^\mu\eq\left(x^+,\,x^1,\,x^2,\,x^-\right)$. In terms of
the instant form variables $\underline
x^\mu\eq\left(x^\circ,\,x^1,\,x^2,\,x^3\right)$, we have $x^\pm\eq
x^\circ\,\pm\,x^3$. The evolution of a system is studied along the coordinate
$x^+$, and as such it plays the role of ``time.'' In the
nuclear physics community
the words ``spinors'' and ``relativistic wave functions'' are often used
interchangeably. We will adhere to this usage here. For the more formally
minded reader what we present are  the {\it front form} objects,  and
their relation with the instant form counterparts, in the  representation space
on which the finite dimensional realisations of $SU^L(2)$  and $SU^R(2)$
generated by $\vec J\,\pm\,i\,\vec K$ act.

\section{FRONT FORM HADRONIC FIELDS}

We will work under the assumption that the center of mass of a composite
hadrons is best described by the $(j,0)\oplus(0,j)$ wave functions in the front
form. These wave functions will be constructed from the right handed,
$\phi^{R}(p^\mu)$, and left handed, $\phi^{L}(p^\mu)$, matter fields.
We begin  form transformation which takes a particle
from rest, $\po^\mu\eq(m,\,0,\,0,\,m)$,
to a particle moving with an arbitrary four momentum $p^\mu\eq
\left(p^+,\,p^1,\,p^2,\,p^-\right)$
[Note: for massive particles $p^+ \,>\, 0$.].

In the {\it instant form of field theory} the transformation
which takes the rest momentum
${\underline\po}^\mu\eq \left(m,\,0,\,0,\,0\right)\,\to\,
\underline {p}^\mu\eq(p^\circ,\,\vec p\,)$,
is constructed out of the boost operator $\vec K$ and is given by
\beq
{\underline p}^\mu \eq{\Lambda^\mu}_\nu\,\underline\po^\nu\,,
 \label{i}
\eeq
with
\beq
\Lambda\eq\exp\left(i\vec \varphi\cdot\vec K\right)\,\,.\label{il} \eeq
In Eq. (\ref{il}) the boost parameter $\vec \varphi$ is defined as follows
\beq
\hat\varphi\eq{\vec p/|\vec p\,|},~~\cosh(\varphi)\eq E/m,~~\sinh(\varphi)\eq
|\vec p\,|/m.\label{ibp}
\eeq
Note that the stability group of the $x^\circ\eq 0$ plane consists of the six
generators $\vec J$ and $\vec P$. The $\vec K$, along with the $P^\circ$,
generate the instant-form dynamics.

In the {\it front form of field theory} the transformation which takes
${\underline\po}^\mu\eq \left(m,\,0,\,0,\,0\right)\,\to\,
\underline p^\mu\eq(p^\circ,\,\vec p\,)$
is given \cite{Soper,LeutwylerS}  by
\beq
\underline{p}^\mu\eq {L^\mu}_\nu\,\underline{\po}^\nu, \label{onee}
\eeq
with the matrix $L$ given by
\beq
L\eq\exp\left(i\,\vec v_\perp\cdot\vec G_\perp\right)
\,\exp\left(i\,\eta K_3\right). \label{boost}
\eeq

The parameters  $\eta$ and $\vec v_\perp = (v_x, v_y)$ specify a given boost.
The  generators $\vec G_\perp$,
which along with  $ K_3$,  take $\underline{\po}^\mu\,\to\,\underline{p}^\mu$,
are defined as follows
\beq
G_1\eq K_1\,-\,J_2,~~G_2\eq K_2\,+\,J_1,\label{g}
\eeq
and together with
\beq
P_- \eq P_\circ\, -\, P_3,~~P_1,~~P_2,~~J_3,
\eeq
form the seven generators of the stability group of the
 $x^+\eq 0$ plane. (Note that $P_- = P^+$).
The algebra associated with the stability group is summarised in Table I.
The generators $D_1\eq K_1\,+\,J_2$, $D_2\eq K_2\,-\,J_1$ and $P_+\eq
P_\circ \,+\, P_3$ generate the front-form dynamics.

It is important to note that while the front-form
transformation $L$ is specified
entirely in terms of the generators of the $x^+\eq 0$ plane stability group,
the
instant-form transformation $\Lambda$ involves dynamical generators associated
with the $x^\circ\eq 0$ plane.

Using the explicit matrix expressions for $\vec J = (J_{1},J_{2},J_{3})$ and
$\vec K = (K_{1},K_{2},K_{3})$ given in
Eqs. (2.65 - 2.67) of Ref.
\cite{Ryder} we obtain an explicit expression for
the  boost $L$ defined in Eq. (\ref{boost})
\widetext
\begin{eqnarray}
[{L^\mu}_\nu]\eq
\left[
\begin{array}{cccccc}
\cosh(\eta)+{1\over 2} \vec v_\perp^{\,2} \exp(\eta)& & v_x & v_y &&
\sinh(\eta)+{1\over 2} \vec v_\perp^{\,2} \exp(\eta)\\
v_x \exp(\eta)& & 1 & 0 &&v_x \exp(\eta)\\
v_y \exp(\eta) && 0 & 1 &&v_y \exp(\eta)\\
\sinh(\eta)-{1\over 2} \vec v_\perp^{\,2} \exp(\eta) && v_x & v_y &&
\cosh(\eta)-{1\over 2} \vec v_\perp^{\,2} \exp(\eta)\\
\end{array}
\right]
. \label{l}
\end{eqnarray}

Recalling that the components of the front form momentum $p^\mu$ are defined as
$p^\pm\eq p^\circ\,\pm\,p^3$ this yields

\widetext
\beq
p^+\eq m\,\exp(\eta),~~\vec p_\perp\eq m\,\exp(\eta)\,\vec v_\perp,~~p^-
\eq m\,\exp(-\eta)\,+\,m\,\exp(\eta) \,\vec v_\perp^{\,2}\,\,.\label{pp}
\eeq

The variables $\eta$ and  $\vec v_\perp$  are fixed by requiring
$\underline {p}^\mu$ generated by Eq. (1) to be identical to the $\underline
{p}^\mu$ produced by Eq. (4).

Given the transformation $L$, Eq. (\ref{boost}), we now wish to construct
the front form
$(j,0)\oplus(0,j)$ hadronic wave functions. To proceed in this direction we
rewrite $L$ by expanding the exponentials in Eq. (\ref{boost}),
and using Table I to to arrive at \cite{msBHF}
\beq
L\eq\exp\left[i\left(a\, \vec v_\perp\cdot\vec G_\perp\,+\,\eta\,
 K_3\right)\right],
 \label{mikolaj}
\eeq
with
\beq
a\eq{\eta\over{1\,-\,\exp(-\eta)}}.\label{mikolaja}
\eeq
For the $(j,0)$ matter fields, $\phi^R(p^\mu)$, we have
\cite{Weinberg,Ryder,dva,mpe}  $\vec K\eq -\,i\,\vec J$. For the $(0,j)$ matter
fields, $\phi^L(p^\mu)$,  $\vec K\eq +\,i\,\vec J$.  Using this observation,
along with Eq. (\ref{mikolaj}) and definitions (\ref{g}), we obtain the
transformation properties of the front form $(j,0)$ and $(0,j)$ hadronic fields
\beq
\phi^R(p^\mu)\eq\exp\left(+\,\eta \,\hat b\cdot\vec
J\,\right)\,\phi^R(\po^\mu),
\label{right}
\eeq
and
\beq
\phi^L(p^\mu)\eq\exp\left(-\,\eta\, \hat b^\ast\cdot\vec J\,\right)\,
\phi^R(\po^\mu).
\label{left}
\eeq
In Eqs. (\ref{right}) and (\ref{left}) we have introduced
 the  unit vectors $\hat b$
and its complex conjugate $\hat b^\ast$ as follows
\begin{eqnarray}
\hat b&\eq&\eta^{-1}\, \left(a\,v_r\,,\,\,-i\, a\, v_r\,,\,\,\eta\right)\, ,
\label{b}
\\
\hat b^\ast&\eq& \eta^{-1}\,\left(a\,v_\ell\,,\,\,i\, a\, v_\ell\,,\,\,
\eta\right)\,;~~\hat b\cdot
\hat b\eq 1\eq \hat b^\ast\cdot\hat b^\ast\,,\label{bs}
\end{eqnarray}
with $v_r\,\eq\,v_x\,+\,i\,v_y$ and
$v_\ell\,\eq\,v_x\,-\,i\,v_y$.

We now make two observations. First,  under the operation of {\it parity}
\beq
{\rm Parity:}~~~\phi^R(p^\mu)\,\leftrightarrow\, \phi^L(p^\mu).
\label{parity}
\eeq
Second, because of the isotropy of the null vector $\vec p\eq\vec 0$ (As argued
in Ref. \cite{Ryder} and discussed in  more detail in Ref. \cite{prcb})
the concept of handedness looses its physical significance for
$p^\mu\eq\po^\mu$, and this in turn yields  the relation
\beq
\phi^R(\po^\mu)\eq\pm\,\phi^L(\po^\mu).\label{pm}
\eeq
To exploit  these observations we now introduce the spin-j hadronic wave
functions
\beq
\psi(p^\mu)\eq
\left[
\begin{array}{c}
\phi^R(p^\mu)\,+\,\phi^L(p^\mu)\\ \\
\phi^R(p^\mu)\,-\,\phi^L(p^\mu)
\end{array}
\right].\label{psi}
\eeq
The  plus (minus) sign in Eq. (\ref{pm}) yields hadronic wave functions with
{\it
even (odd) intrinsic parity}. We will denote the
even intrinsic parity wave functions by ${\cal U}(p^\mu)$; and the odd
intrinsic parity wave functions by
${\cal V}(p^\mu)$.

The transformation property  for these hadronic wave functions under the boost
(\ref{boost}) is now readily obtained by using  Eqs. (\ref{right}) and
(\ref{left}). The result is
\beq
\psi(p^\mu)\eq M(L)\,\,\psi(\po^\mu)\,,\label{tran}
\eeq
with the hadronic-wave-function boost operator, $M(L)$,  given by
\widetext
\beq
M(L)\eq\left[
\begin{array}{ccc}
\exp\left(\eta\,\hat b\cdot\vec J\,\right)
\,+\, \exp\left(-\,\eta\,\hat b^\ast\cdot\vec J\,\right)&~~~~~~~&
\exp\left(\eta\,\hat b\cdot\vec J\,\right)
\,-\, \exp\left(-\,\eta\,\hat b^\ast\cdot\vec J\,\right)\\ \\
\exp\left(\eta\,\hat b\cdot\vec J\,\right)
\,-\, \exp\left(-\,\eta\,\hat b^\ast\cdot\vec J\,\right)&~~~~~~~&
\exp\left(\eta\,\hat b\cdot\vec J\,\right)
\,+\, \exp\left(-\,\eta\,\hat b^\ast\cdot\vec J\,\right)
\end{array}
\right].\label{ml}
\eeq
The construction of the hadronic wave functions ${\cal U}(p^\mu)$ and ${\cal
V}(p^\mu)$, introduced above, for a specific spin requires: {\it (i)}. Explicit
evaluation of the $2(2j\,+\,1)\times 2(2j\,+\,1)$ matrix $M(L)$ defined via Eq.
(\ref{ml}); and {\it (ii)}. Some other necessary technical details (see below).
We
establish this procedure in the next section by explicitly constructing ${\cal
U}(p^\mu)$ and ${\cal V}(p^\mu)$ for $j\eq{1\over 2},\,1,{3\over 2},\,2$.

\section{Construction of ${\cal U}(\lowercase {p}^\mu)$ and ${\cal V}
(\lowercase{p}^\mu)$ and Their Properties}

For obtaining  specific explicit expressions for the hadronic wave functions
${\cal U}(p^\mu)$ and ${\cal V}(p^\mu)$, the technical detail which still needs
to be taken care of is to fix the representation  for the generators of
rotation $\vec J$. Towards this end we note that the front form helicity
operator
\beq
{\cal J}_3 \equiv J_3 \,+\,{1\over P_-}\left(G_1\,P_2\,-\,G_2\,P_1\right),
\label{o}
\eeq
introduced by Soper \cite{Soper} and discussed by Leutwyler
and Stern \cite{LeutwylerS}, commutes with all generators of the stability
group
associated with the $x^+\eq0$ plane. The front form helicity  operator
associated with the $(j,0)\oplus(0,j)$ hadronic fields constructed above is
then readily defined to be
\beq
\Theta\eq\left[
\begin{array}{ccc}
{\cal J}_3 &~~& 0\\
0 &~~&{\cal J}_3
\end{array}\right]\,\,. \label{th}
\eeq
Taking a matrix representation of the $\vec J$ operators with $J^3$ diagonal
(in the standard convention of Ref. \cite{Schiff}), the
$2(2j\,+\,1)$--element basis spinors for the hadronic wave functions
which correspond to $p^\mu\eq\po^\mu$ have then the general form
\widetext
\beq
{\cal U}_{+j}(\po^\mu)\eq
\left[
\begin{array}{c}
{\cal N}(j)\\
0\\
0\\
\vdots\\
0
\end{array}
\right],~~
{\cal U}_{j-1}(\po^\mu)\eq
\left[
\begin{array}{c}
0\\
{\cal N}(j)\\
0\\
\vdots\\
0
\end{array}
\right],\ldots\ldots,\,
{\cal V}_{-j}(\po^\mu)\eq
\left[
\begin{array}{c}
0\\
0\\
0\\
\vdots\\
{\cal N}(j)
\end{array}
\right]\,\,.\label{basis}
\eeq
The index $h\eq j,\, j-1,\,\ldots,\,-j$ on the ${\cal U}_h(p^\mu)$ and ${\cal
V}_h(p^\mu)$ corresponds to the eigenvalues of the the front form helicity
operator $\Theta$. The spin-dependent normalisation constant ${\cal N}(j)$ is
to be so chosen that for the {\it massless} particles  the ${\cal
U}_h(\po^\mu)$ and ${\cal V}_h(\po^\mu)$ vanish (There can be no massless
particles at rest !); and only the ${\cal U}_{h\eq\pm j}(p^\mu)$ and ${\cal
V}_{h\eq\pm j}(p^\mu)$ survive in the high energy limit. As will be shortly
verified the simplest choice satisfying these requirements is
\beq
{\cal N}(j)\eq m^{\,j}.\label{nj}\label{norm}
\eeq

We now have all the details needed to construct ${\cal U}_h(p^\mu)$
and ${\cal V}_h(p^\mu)$ for any hadronic field. In this paragraph we
summarise the algebraic construction used for $j\eq{1\over
2},\,1,{3\over 2},\,2$. Using Eqs. (A27,A28) and (A31,A32) of Ref.
\cite{Weinberg} along with Eqs.  (\ref{pp},\ref{b},\ref{bs}) above we
obtain the the expansions for the $\exp\left(\eta\,\hat b\cdot\vec
J\,\right)$ and $ \exp\left(-\,\eta\,\hat b^\ast\cdot\vec J\,\right)$
which appear in the hadronic-wave-function boost matrix $M(L)$, Eq.
(\ref{ml}). These expansions are presented in Appendix A.  Using the
results of Appendix A, explicit expressions for $M(L)$, Eq.
(\ref{ml}), are then calculated as a simple, but somewhat lengthy,
algebraic exercise. The $M(L)$ so obtained in conjunction with Eqs.
(\ref{tran}) and (\ref{basis}), then yield the hadronic wave functions
presented in Appendix B. The generality of the procedure for any spin
is now obvious, and the procedure reduces to a well defined algebraic
manipulations.

We now introduce the following useful matrices,
\beq
\Gamma^\circ\eq
\left[
\begin{array}{ccc}
I&{~}&0\\
0&{~}&-I
\end{array}
\right]\,,
{}~~
\Gamma^{5}\eq
\left[
\begin{array}{ccc}
0&{~~}&I \\
I&{~~}& 0
\end{array}
\right]\, ,
\label{go}
\eeq
with $I\eq (2j\,+\,1)\times(2j\,+\,1)$ identity matrix.

In reference to the hadronic wave functions presented in Appendix B, we define
\beq
\overline\psi\l p^\mu \r \equiv \psi^\dagger\l p^\mu\r\,\Gamma^\circ\, .
\label{psibar}
\eeq

Using the explicit
expressions for ${\cal U}(p^\mu)$ and ${\cal V}(p^\mu)$, Eqs.
(B1-B7), we verify that
\begin{eqnarray}
{\overline {\cal U}}_h (p^\mu)\,\,{\cal U}_{h'}(p^\mu) &\eq&
m^{2\,j}\,\delta_{h h'} \,,\label{three}\\
{\overline {\cal V}}_h (p^\mu)\,\,{\cal V}_{h'}(p^\mu) &\eq&
-\,m^{2\,j}\,\delta_{h h'} \,,\label{four}\\
{\overline {\cal U}}_h (p^\mu)\,\,{\cal V}_{h'}(p^\mu) &\eq& 0
\eq{\overline {\cal V}}_h (p^\mu)\,\,{\cal U}_{h'}(p^\mu)  \,\,.\label{five}
\end{eqnarray}
One of the  most noteworthy feature of the ${\cal U}_h(p^\mu)$ and ${\cal
V}_h(p^\mu)$ constructed here in the $(j,\,0)\oplus(0,\,j)$ representation is
the  observation [28,  Sec. III] that  the high energy limit,
$p^+\,\gg\,m$, is equivalent to the massless case. This observation is
consistent with the explicit results presented in Appendix B: ${\cal
U}_h(p^\mu)$ and ${\cal V}_h(p^\mu)$ {\it identically
vanish in this limit unless the associated front form helicity $h\eq\pm\,j$.}
This is
precisely the type of behaviour that a theory suited for the high energy
phenomenon is expected to have. An examination of Eqs. (B3), (B4) and (B5-B7)
further suggests that  in the high energy limit the  ${\cal U}_h(p^\mu)$
and ${\cal V}_h(p^\mu)$ fall off as $\sim\,\l m/\pp\r ^{j\,-\,|h|}$.

Finally, we note that for spin one half the result given by Eq. (B1) is
identical to that given by Lepage and Brodsky [19, Eq.A3]. Note however a
slightly different normalisation and their choice of phase for the odd
intrinsic parity wave functions.

\section{The Connection with the Instant Form}

In a representation appropriate for comparison with the front form
spinors ${\cal U}_h(\lowercase {p}^\mu)$ and ${\cal
V}_h(\lowercase{p}^\mu)$, the instant form hadronic wave functions
$u_\sigma(\,{\underline p}^\mu)$ and $v_\sigma (\,{\underline
p}^\mu)$, $\sigma\eq j,\,j\,-\,1,\ldots,-\,j$, were recently
constructed (following Weinberg \cite{Weinberg} and Ryder\cite{Ryder})
explicitly in Refs. \cite{dva,prc,mpe}. A brief report, sufficient for
the present discussion, can be found in Ref. \cite{prc}. Here we only
remark that the construction of instant form spinors follows the steps
outlined in Eqs.(\ref{mikolaj}-\ref{ml}) above, with the only difference
that one starts with transformation $\Lambda$ of Eq.(\ref{il}) rather
than $L$ of Eq.(\ref{mikolaj}).

The instant form hadronic wave functions
of Refs.
\cite{dva,prc,mpe} satisfy the following normalisation properties
\begin{eqnarray}
{\overline u}_\sigma (\,\underline {p}^\mu)\,\,{u}_{\sigma'}
(\underline{p}^\mu) &\eq&
m^{2\,j}\,\delta_{\sigma \sigma'} \\
{\overline v}_h (\,\underline{p}^\mu)\,\,v_{\sigma'}(\,\underline{p}^\mu) &\eq&
-\,m^{2\,j}\,\delta_{\sigma\sigma'} \\
{\overline u}_h (\,\underline{p}^\mu)\,\,{v}_{h'}(\,\underline{p}^\mu) &\eq&
0\eq
{\overline v}_h (\,\underline{p}^\mu)\,\,{u}_{h'}(\,\underline{p}^\mu) \,\,.
\end{eqnarray}
where
\beq
\overline{u}_\sigma(\,\underline{p}^\mu) \eq
\left[{u}_\sigma(\,\underline{p}^\mu)
\right]^\dagger\gamma^\circ,~~
\overline{v}_\sigma(\,\underline{p}^\mu) \eq
\left[{v}_\sigma(\,\underline{p}^\mu)
\right]^\dagger\gamma^\circ
\eeq
with $\gamma^\circ\eq$ having the form identical to $\Gamma^\circ$ of
Eq.(\ref{go}). In what follows we assume that $p^\mu$ and $\underline{p}^\mu$
correspond to the same physical momentum.

The connection between the front form and instant form is established
by noting that on general algebraic grounds, we can express the instant
form hadronic wave functions as linear combination of the front form
hadronic wave functions. That is
\begin{eqnarray}
u_\sigma( \,\underline{p}^\mu)&\eq& \Omega^{(u\,{\cal U})}_{\sigma h}\,{\cal
U}_h( p^\mu) \,+\, \Omega^{(u\,{\cal V})}_{\sigma h}\,{\cal V}_h(p^\mu) \,,
\label{one} \\
v_\sigma( \,\underline{p}^\mu)&\eq& \Omega^{(v\,{\cal U})}_{\sigma h}\,{\cal
U}_h( p^\mu) \,+\, \Omega^{(v\,{\cal V})}_{\sigma h}\,{\cal V}_h(p^\mu) \,,
\label{two}
\end{eqnarray}
where the sum on the repeated indices is implicit.

We now multiply Eq. (\ref{one}) by ${\overline{\cal U}}_{h'}(p^\mu)$ from the
left,
and using the orthonormality relations, Eqs. (\ref{three}-\ref{five}),
we get
\beq
\Omega^{(u\,{\cal U})}_{\sigma h} \eq {1\over m^{2j}}\left[\,{\overline{\cal
U}}_h
(p^\mu)\,
u_\sigma(\,\underline{p}^\mu)\right]\,\,. \eeq
Similarly by multiplying Eq. (\ref{two}) from the left by
${\overline{\cal V}}_{h'}(p^\mu)$ and again using
the orthonormality relations, Eqs. (\ref{three}-\ref{five}), we obtain
\beq
\Omega^{(u\,{\cal U})}_{\sigma h} \eq {1\over m^{2j}}\left[\,
{\overline{\cal U}}_h
(p^\mu)\,
u_\sigma(\,\underline{p}^\mu)\right]\,\,.
\eeq
Further it is readily verified, e.g. by using the results presented in Appendix
B here and explicit expressions for $u_\sigma(\,\underline{p}^\mu)$ and
$v_\sigma(\,\underline{p}^\mu)$ found in Refs. \cite{dva,prc,mpe}, that
\beq
{\overline{\cal V}}_h(p^\mu)\,u_\sigma(\,\underline{p}^\mu)\eq 0
\eq
{\overline{\cal U}}_h(p^\mu)\,v_\sigma(\,\underline{p}^\mu);
\eeq
which yields
\beq
\Omega^{(u\,{\cal V})}_{\sigma h} \eq 0 \eq
\Omega^{(v\,{\cal U})}_{\sigma h}\,\, .
\eeq
Finally, we exploit the facts
\beq
\left\{\Gamma^5,\,\Gamma^\circ\right\}\eq 0,~~{\Gamma^5}^\dagger \eq
\Gamma^5,~~ {\rm and}~\left(\Gamma^5\right)^2\eq I,
\eeq
to conclude that ${\overline {\cal V}}_h( p^\mu)\,v_\sigma(\,\underline{p}^\mu)
\eq -\, {\overline {\cal U}}_h( p^\mu)\,u_\sigma(\,\underline{p}^\mu)$. Thus
defining a $(2j\,+\,1) \times(2j\,+\,1)$ matrix $B(j)$ such that $B_{\sigma
h}\eq \overline{\cal U}_h(p^\mu)\,u_\sigma(\,\underline{p}^\mu)\eq
\Omega^{(u{\cal U})}_{\sigma h} \eq
\Omega^{(v{\cal V})}_{\sigma h}
$ we get the matrix which connects the instant form hadronic wave functions
with the front form wave functions. It reads
\beq
\Omega(j)\eq
\left[
\begin{array}{ccc}
B(j)&{~}& 0 \\ \\
0 &{~}& B(j)
\end{array}
\right]\,\,.\label{six}
\eeq

The explicit expressions for $\Omega(j)$ are presented in Appendix C.
For spin one half the transformation matrix $\Omega(1/2)$ computed by
us coincides with the celebrated ``Melosh transformation'' given by
Melosh in Ref. [20, Eq.  26] and by Dziembowski in Ref.  [21, Eq. A8].
As formally demonstrated by Kondratyuk and Terent'ev \cite{KT}, the
transformation matrix $\Omega({j})$ represents a pure rotation of the
spin basis. {\it However, since $\Omega(j)$ has block zeros off-diagonal,
what manifestly emerges here is that this rotation does not mix the even
and odd intrinsic parity wave functions.}

\section{Summary and Concluding Remarks}

We begin with a summary of the results obtained. In this paper we have
presented a relativistic formalism to construct hadronic fields of
arbitrary spin in the front form, and established their connection
with the instant form fields. For a given a spin, $j$, there are
$(2j\,+\,1)$ hadronic wave functions with {\it even} intrinsic parity,
${\cal U}_h(p^\mu)$, and $(2j\,+\,1)$ hadronic wave functions with
{\it odd} intrinsic parity, ${\cal V}_h(p^\mu)$.  From a formal point
of view, the fundamental object required to construct these wave
functions is the hadronic-wave-function boost matrix $M(L) $ in the
front form. The matrix $M(L) $ is presented in Eq.  ({\ref{ml}). The
explicit construction of the front form $(j,\,0)\oplus(0,\,j)$
hadronic spinors requires the introduction of the front form helicity
operator ${\cal J}_3$, Eq. (\ref{o}), introduced by Soper \cite{Soper}
and, Leutwyler and Stern \cite{LeutwylerS}. Further,  the normalisation of
these wave functions should be so chosen that for the {\it massless} particles
the ${\cal U}_h(\po^\mu)$ and ${\cal V}_h(\po^\mu)$  identically vanish; and
only ${\cal U}_{h\eq\pm j}(p^\mu)$ and ${\cal V}_{h\eq\pm j}(p^\mu)$ survive in
the high energy limit. The simplest choice of this normalisation is given
by Eq. (\ref{norm}). Next we constructed a matrix $\Omega(j)$ which provides
the connection between the front form hadronic wave functions with the more
familiar [i.e. more ``familiar'' at least for spin one half case] instant form
hadronic wave functions. We discovered that  the transformation matrix
$\Omega(1/2)$ coincides with the well known ``Melosh transformation''
\cite{Melosh,Dziembowski}, and the spin one half wave functions are in
agreement with the previous results of Lepage and Brodsky \cite{lepage}.
Explicit results for ${\cal U}_h(p^\mu)$, ${\cal V}_h(p^\mu)$ and $\Omega(j)$
up to spin two are found in Appendixes B and C here.

Having provided a brief summary of the main results of this work, we now take
liberty of making some concluding remarks. While, in principle, the  effective
action of the QCD, $S[\pi,\,\rho,\,\omega,\ldots,N,\,\overline
N,\,\Delta^{3/2},\ldots]$, contains arbitrarily high spins, an effective theory
wishing to describe high energy phenomenon up to a certain energy scale can
comfortably truncate the expansion at a certain physically determined value of
spin. This physical value may be determined, for example,  by the usual partial
wave analysis of the cross sections. Even though  the details of this procedure
and the actual connecting link with QCD does not yet exist in detail,  there is
little doubt that the hadronic fields constructed here are one of the most
natural objects in terms of which to express a phenomenological and effective
action of medium and high energy hadronic and nuclear physics. For perturbative
calculations we can construct front form field operators $\Psi[x^\mu,\,j]$ as
expansions in terms of the front form fields ${\cal U}_h(p^\mu)$ and ${\cal
V}_h(p^\mu)$ presented here, and then evaluate the vacuum expectation value of
the $x^+$ ordered product $\langle~~| X^+\,\left[\Psi[x^\mu,\,j]\,
\overline{\Psi}[y^\mu,\,j]\,\right]|~~\rangle$ to obtain the front form
Feynman-Dyson  propagators. The $x^+$ ordered product of a set of arbitrary
front form field operators $\Psi_{{(\alpha)}}[x^\mu_{{(\alpha)}}]\,
\Psi_{(\beta)}[x^\mu_{(\beta)}]\,\ldots\,
\Psi_{(\epsilon)}[x^\mu_{(\epsilon)}]$ is defined as follows
\beq
X^+\left[
\Psi_{(\alpha)}[x^\mu_{(\alpha)}]\,
\Psi_{(\beta)}[x^\mu_{(\beta)}]\ldots
\Psi_{(\epsilon)}[x^\mu_{(\epsilon)}] \right] \eq
(-1)^f\,
\Psi_{(\omega)}[x^\mu_{(\omega)}]\,
\Psi_{(\lambda)}[x^\mu_{(\lambda)}]\ldots
\Psi_{(\rho)}[x^\mu_{(\rho)}] \,\,,
\eeq
where on the right hand side, the field operators are the same ones
as on the left but are ordered in such a order that
\beq
x^+_{(\omega)}\, \ge\, x^+_{(\lambda)}\,\ge\, \ldots \,\ge\,x^+_{(\rho)}\, ,
\eeq
and $f\eq$ the number of necessary transpositions among the fermion field
operators that are needed to achieve the ordering. These $x^\mu$-space
propagators can be Fourier transformed to obtain the $p^\mu$-space front form
Feynman-Dyson propagators. At this stage one can either further
 develop  a front form
formalism similar to Weinberg's equations in the instant form to
guide one self to define a class of relativistic phenomenologies
characterised by a specific choice of
$S[\pi,\,\rho,\,\omega,\ldots,N,\,\overline N,\,\Delta^{3/2},\ldots]$,
or invert the
$p^\mu$-space front form Feynman-Dyson propagator as recently done in Ref.
\cite{plb} to obtain  a class of non-local (for $j\,>\, 1/2$; for the $j\eq1/2$
the equation is identical to the Dirac equation)  wave equations
in the instant form.
Even though it
 has been   recently shown that Weinberg equations for arbitrary spin
suffer from
certain kinematically spurious \cite{prc,wig} solutions, we
have been able to develop \cite{mikkel}
a Lorentz covariant procedure which prevents the interactions to induce
transitions between the kinematically acceptable solutions (which are
the counterpart of the front form wave functions in the instant form) and
kinematically unacceptable solutions.

Apart from these practical aspects, the formalism has certain in-built elegance
and exhibits great beauty of the front form of field theory. The work presented
here, as already indicated, provides the unifying link between the early works
of Weinberg \cite{Weinberg} and Dirac \cite{DiracSpin} on high spin fields
  and their
later works on infinite momentum frame \cite{WeinbergI} and front form
\cite{DiracLight} of field theory.

\acknowledgements

DVA extends his warmest thanks to Ovid C. Jacob, Mikkel B. Johnson and R. M.
Thaler for insightful conversations on the general subject of this work.
He  also acknowledges a great indebetedness to David J. Ernst for our previous
collaborative work on this subject, and the insights which are carried over
into the present work. {\it Zimpoic} thanks are due to Iris Wellner and
Christoph Burgard.  DVA also thankfully acknowledges financial support via a
postdoctoral fellowship by the Los Alamos National Laboratory. Part of this
work was  done while one of the authors (M.S.) was a visiting faculty at the
Department of Physics of the Texas A\&M University. He would like to gratefully
thank Professor David J. Ernst and  members of the Department of Physics for
warm hospitality extended to him during his stay. The financial support of the
NSF under the Grant PHY-8907852 is gratefully acknowledged. \newpage

\widetext
\appendix{Expansions for the $\exp\left(\eta\,\lowercase{\hat b}\cdot\vec
J\,\right)$ and $ \exp\left(-\,\eta\,\lowercase{\hat b}^\ast\cdot\vec
J,\right)$ up to Spin Two}
\footnotesize
\widetext
This appendix provides the expansions for the $\exp\left(\eta\,\hat b\cdot\vec
J\,\right)$ and $ \exp\left(-\,\eta\,\hat b^\ast\cdot\vec J\,\right)$, up to
$j\eq 2$,  which appear in the hadronic-wave-function boost matrix $M(L)$, Eq.
(\ref{ml}).

\centerline{ $j\eq{1\over 2}$ }
\begin{eqnarray}
\exp\left(\eta\,\hat b\cdot\vec J\,\right)
&\eq&
\cosh\left({\eta/ 2}\right) \,I\,+\,
\left(\hat b\cdot\vec\sigma\right)\,\sinh\left({\eta/ 2}\right)\,, \label{aa}
\\
\exp\left(-\,\eta\,\hat b^\ast\cdot\vec J\,\right)
&\eq&
\cosh\left({\eta/2}\right)\,I \,-\,
\left(\hat b^\ast\cdot\vec\sigma\right)\,\sinh\left({\eta/ 2}\right)\,.
\label{ab}
\end{eqnarray}
\def\bj{\left(\hat b\cdot\vec J\,\right)}
\def\bsj{\left(\hat b^\ast\cdot\vec J\,\right)}
\centerline {$j\eq{1}$ }
\begin{eqnarray}
\exp\left(\eta\,\hat b\cdot\vec J\,\right)
&\eq&
I\,+\,2\bj^2\sinh^2(\eta/2)\,+\,2\bj\cosh(\eta/2)\,\sinh(\eta/2)\,,\\
\exp\left(-\,\eta\,\hat b^\ast\cdot\vec J\,\right)
&\eq&
I\,+\,2\bsj^2\sinh^2(\eta/2)\,-\,2\bsj\cosh(\eta/2)\,\sinh(\eta/2)\,.
\end{eqnarray}

\def\bbj{\left(2\hat b\cdot\vec J\,\right)}
\def\bbsj{\left(2\hat b^\ast\cdot\vec J\,\right)}
\def\e{(\eta/2)}
\centerline{ $j\eq{3\over 2}$ }
\begin{eqnarray}
\exp\left(\eta\,\hat b\cdot\vec J\,\right)
\eq&&
\cosh\e\left[I\,+\,{1\over 2}\left\{\bbj^2\,-\,I\right\}\,\sinh^2\e\right]
\nonumber\\
&&{}\,+\,\bbj\sinh\e\left[I\,+\,{1\over 6}\left\{\bbj^2\,-\,I\right\}\sinh^2\e
\right] \,,\\
\exp\left(-\,\eta\,\hat b^\ast\cdot\vec J\,\right)
\eq&&
\cosh\e\left[I\,+\,{1\over 2}\left\{\bbsj^2\,-\,I\right\}\,\sinh^2\e\right]
\nonumber\\
&&{}\,-\,\bbsj\sinh\e\left[I\,+\,{1\over
6}\left\{\bbsj^2\,-\,I\right\}\sinh^2\e
\right]\,.
\end{eqnarray}

\centerline{ $j\eq 2$ }
{\footnotesize
\begin{eqnarray}\hskip-.5in
\exp\left(\eta\,\hat b\cdot\vec J\,\right)
\eq&&
I\,+\,2\bj^2\sinh^2\e\,+\,{2\over 3}\bj^2\left\{\bj^2\,-I\right\}\sinh^4\e
 \nonumber\\
&&{}\,+\, 2\bj\cosh\e\sinh\e\,+\,{4\over
3}\bj\left\{\bj^2\,-\,I\right\}\cosh\e\sinh ^3\e\,,\\
\exp\left(-\eta\,\hat b^\ast\cdot\vec J\,\right)
\eq&&
I\,+\,2\bsj^2\sinh^2\e\,+\,{2\over 3}\bsj^2\left\{\bsj^2\,-I\right\}
\sinh^4\e \nonumber \\
&&{}\,-\, 2\bsj\cosh\e\sinh\e\,-\,{4\over 3}\bsj\left\{\bsj^2\,-\,I\right\}
\cosh\e\sinh
^3\e\,.
\end{eqnarray}
}
In Eqs. (\ref{aa},\ref{ab}) the $\vec \sigma$ are the standard
\cite{Ryder} Pauli matrices.
In this appendix, $I$ are  the $(2j+1)\times(2j+1)$ identity matrices.

\appendix{Front Form Hadronic Wave Functions up to Spin two}
\footnotesize
We begin with  collecting together front form hadronic wave functions up to
spin 2 {\it for even intrinsic parity}, first. In what follows we use the
notation
$\pr\eq p_x\,+\,i\,p_y$ and $\pl\eq p_x\,-\,i\,p_y$, c.f. Eq.(\ref{bs}),
\vskip 0.3in\noindent
Spin one half hadronic wave functions with
even intrinsic parity:

\def\ca{{1\over{2}}\sqrt{1\over \pp}}
\beq
{\cal U}_{+\,{1\over 2}}(p^\mu)\eq\ca\,\left[
\begin{array}{c}
\pp\,+\,m\\ \\
\pr\\ \\
\pp\,-\,m\\ \\
\pr
\end{array}
\right]
\,,~~
{\cal U}_{-\,{1\over 2}}(p^\mu)\eq\ca\left[
\begin{array}{c}
-\,\pl\\ \\
\pp\,+\,m\\ \\
\pl\\ \\
-\,\pp\,+\,m
\end{array}
\right]
\,.\label{ua}
\eeq

\vskip 0.3in\noindent
Spin one  hadronic wave functions with
even intrinsic parity:
{\footnotesize\hskip-.5in
\beq
{\cal U}_{+1}(p^\mu)\eq
{1\over 2}\,\left[
\begin{array}{c}
\pp\,+\,\left(m^2/\pp\right)\\ \\
{\sqrt{2}}\,\pr \\ \\
{\pr^2/\pp}\\ \\
\pp\,-\,\left({m^2/\pp} \right)\\ \\
{\sqrt{2}} \,\pr\\ \\
{\pr^2/\,\pp}
\end{array}
\right],\,~~
{\cal U}_{0}(p^\mu)\eq
m\,{\sqrt{1\over 2}}\,\left[
\begin{array}{c}
-\,{\pl/\pp} \\\\
\sqrt {2} \\ \\
\,{\pr/\pp} \\ \\
\,{\pl/\pp} \\\\
0\\ \\
\,{\pr/\pp}
\end{array}
\right]\,~~
{\cal U}_{-1}(p^\mu)\eq
{1\over 2}\,\left[
\begin{array}{c}
{\pl^2/\pp}  \\ \\
-\,{\sqrt{2}} \,\pl\\ \\
\pp\,+\,\left({m^2/\pp} \right)\\ \\
-\,{\pl^2/\pp}  \\ \\
{\sqrt{2}}\, \pl\\ \\
-\,\pp\,+\,\left({m^2/\pp} \right)
\end{array}
\right]\,.\label{ub}
\eeq
}

\def\t{\sqrt{3}}

\vskip 0.3in\noindent
Spin three half  hadronic wave functions with
even intrinsic parity:
\beq
{\cal U}_{+{3\over 2}}(p^\mu)\eq
{{1\over 2}\,\sqrt{1\over\pp}}\left[
\begin{array}{c}
{\pp}^2\,+\,\left(m^3/{\pp} \right) \\ \\
\sqrt{3}\, \pr\,{\pp} \\ \\
\t\,\pr^2 \\ \\
\pr^3/{\pp} \\ \\
{\pp}^2\,-\,\left(m^3/{\pp}\right) \\ \\
\sqrt{3}\, \pr\,{\pp} \\ \\
\t\,\pr^2 \\ \\
\pr^3/{\pp}
\end{array}
\right]\,,~~
{\cal U}_{+{1\over 2}}(p^\mu)\eq
{{m\over 2}\,\sqrt{1\over\pp}}\left[
\begin{array}{c}
-\,\t\,m\,\pl/\pp \\ \\
\pp\, +\,m\\ \\
2\,\pr \\ \\
\t\,{\pr}^2/\pp \\ \\
\t\,m\,\pl/\pp \\ \\
\pp\,-\,m \\ \\
2\,\pr \\ \\
\t\,\pr^2/\pp
\end{array}
\right]\,,\label{uca}
\eeq

\beq
{\cal U}_{-{1\over 2}}(p^\mu)\eq
{{m\over 2}\,\sqrt{1\over\pp}}\left[
\begin{array}{c}
\t\,\pl^2/\pp \\ \\
-\,2\,\pl \\ \\
\pp\,+\,m \\ \\
\t\,m\,\pr/\pp \\ \\
-\,\t\,\pl^2/\pp \\ \\
2\,\pl \\ \\
-\,\pp\,+\,m \\ \\
\t\,m\,\pr/\pp
\end{array}
\right]\,,~~
{\cal U}_{-{3\over 2}}(p^\mu)\eq
{{1\over 2}\,\sqrt{1\over\pp}}\left[
\begin{array}{c}
-\,\pl^3/\pp \\ \\
\t\,\pl^2 \\ \\
-\,\t\,\pl\,\pp \\ \\
{\pp}^2\,+\,\left(m^3/\pp\right) \\ \\
\pl^3/\pp \\ \\
-\t\,\pl^2 \\ \\
\t\,\pl\,\pp \\ \\
-\,{\pp}^2\,+\,\left(m^3/\pp \right)
\end{array}
\right]\,\,.\label{ucb}
\eeq

\vskip 0.3in\noindent
Spin two hadronic wave functions with
even intrinsic parity:
\def\s{\sqrt{6}}
\beq
{\cal U}_{+{2}}(p^\mu)\eq
{1\over 2}\,\left[
\begin{array}{c}
{{\pp}^2}\,+\,\left(m^4/{\pp}^2\right) \\ \\
2\,\pr\,\pp \\ \\
\s\,\pr^2 \\ \\
2\,{\pr}^3/\pp \\ \\
\pr^4/{\pp}^2 \\ \\
{\pp}^2\,-\,\left(m^4/{\pp}^2\right) \\ \\
2\,\pr\,\pp \\ \\
\s\,\pr^2 \\ \\
2\,\pr^3/\pp \\ \\
\pr^4/{\pp}^2
\end{array}
\right]\,,~~
{\cal U}_{+{1}}(p^\mu)\eq
{m\over 2}\,\left[
\begin{array}{c}
-\,2\,m^2\,\pl/{\pp}^2 \\ \\
\pp\,+\,\left(m^2/\pp\right) \\ \\
\s\,\pr \\ \\
3\,\pr^2/\pp \\ \\
2\,\pr^3/{\pp}^2 \\ \\
2\,m^2\,\pl/{\pp}^2 \\ \\
\pp\,-\,\left(m^2/\pp\right) \\ \\
\s\,\pr \\ \\
3\,\pr^2/\pp \\ \\
2\,\pr^3/{\pp}^2
\end{array}
\right]\,,\label{uda}
\eeq

\beq
{\cal U}_{{0}}(p^\mu)\eq
{m^2\over 2}\,\left[
\begin{array}{c}
\s\,\pl^2/{\pp}^2 \\ \\
-\,\s\,\pl/\pp \\ \\
2 \\ \\
\s\,\pr/\pp \\ \\
\s\,\pr^2/{\pp}^2 \\ \\
-\,\s\,\pl^2/{\pp}^2 \\ \\
\s\,\pl/\pp \\ \\
0 \\ \\
\s\,\pr/\pp \\ \\
\s\,{\pr}^2/{\pp}^2
\end{array}
\right]\,,~~
{\cal U}_{-{1}}(p^\mu)\eq
{m\over 2}\,\left[
\begin{array}{c}
-\,2\,\pl^3/{\pp}^2 \\ \\
3\,\pl^2/\pp \\ \\
-\,\s\,\pl \\ \\
\pp\,+\,\left(m^2/\pp\right) \\ \\
2\,m^2\,\pr/{\pp}^2 \\ \\
2\,\pl^3/{\pp}^2 \\ \\
-\,3\,\pl^2/\pp \\ \\
\s\,\pl \\ \\
-\,\pp\,+\,\left(m^2/\pp\right) \\ \\
2\,m^2\,\pr/{\pp}^2 \\ \\
\end{array}\right]\, ,\label{udb}
\eeq

\beq
{\cal U}_{-{2}}(p^\mu)\eq
{1\over 2}\,\left[
\begin{array}{c}
\pl^4/{\pp}^2 \\ \\
-\,2\,\pl^3/\pp \\ \\
\s\,\pl^2 \\ \\
-\,2\,\pl\,\pp \\ \\
{\pp}^2\,+\,\left(m^4/{\pp}^2\right) \\ \\
-\,\pl^4/{\pp}^2 \\ \\
2\,\pl^3/\pp \\ \\
-\,\s\,\pl^2 \\ \\
2\,\pl \,\pp \\ \\
-\,{\pp}^2\,+\,\left(m^4/{\pp}^2\right)
\end{array}
\right]\,\,.\label{udc}
\eeq

An examination of the hadronic-wave-function boost matrix $M(L)$,
Eq.(\ref{ml}), implies that the {\it odd } intrinsic-parity hadronic
wave can be obtained from the hadronic wave functions of the {\it
even} intrinsic parity via the following simple relation
\beq
{\cal V}_h(p^\mu)\eq \Gamma^5 \,{\cal U}_h(p^\mu)\,\,,
\eeq
where the matrix $\Gamma^5$ defined in Eq.(\ref{go}) interchanges the top
$(2j\,+\,1)$ elements with the
bottom $(2j\,+\,1)$ elements of the hadronic wave functions.

\appendix{The Explicit Expressions for $\Omega(\lowercase{j})$ up to Spin Two}
\footnotesize

In this appendix we present explicit expressions for the matrix $\Omega(j)$,
Eq. (\ref{six}),
which connects the front form hadronic wave functions with the hadronic
wave functions via Eqs. (\ref{one},\ref{two}).
As in Appendix B,
$\pr\eq p_x\,+\,i\,p_y$ and $\pl\eq p_x\,-\,i\,p_y$, in what follows.
\vskip 0.3in\noindent

For spin one half,  the matrix connecting the instant form hadronic wave
functions with the front form wave functions is
\def\hc{ {1\over {\left[2\,\l E\,+\,m\r\,\pp\,\right]^{1/2} }}  }
\beq
\Omega\l 1/2 \r \eq \hc \,
\left[
\begin{array}{cccc}
\beta(1/2)&{}&{}&0 \\ \\
0&{}&{}&\beta(1/2)
\end{array}
\right]\,,\label{ha}
\eeq
where the the $2\times2$ block matrix $\beta(1/2)$ is defined as
\beq
\beta\l 1/2\r\eq
\left[
\begin{array}{ccc}
\pp\,+\,m&{}&-\,\pr \\ \\
\pl&{}& \pp\,+\,m
\end{array}
\right]\,.\label{hb}
\eeq


For spin one, the matrix connecting the instant form hadronic wave functions
with the front form wave functions is
\def\oc{ {1\over {\left[2\,\l E\,+\,m\r\,\pp\,\right] }}  }
\beq
\Omega\l 1 \r \eq \oc \,
\left[
\begin{array}{ccc}
\beta(1)&{}&0 \\ \\
0&{}&\beta(1)
\end{array}
\right]\,,\label{oa}
\eeq
where the the $3\times3$ block matrix $\beta(1)$ is defined as
\beq
\beta\l 1 \r \eq
\left[
\begin{array}{ccccc}
\l\pp\,+\,m\r^2 &{}& -\,\sqrt{2}\,\l\pp\,+\,m\r\,\pr &{~~}& \pr^2 \\ \\
\sqrt{2}\,\l\pp\,+\,m\r\,\pl &{~~}&2\,\left[\l E\,+\,m\r\,\pp\,-\,\pr\pl\right]
&{~~}&-\,\sqrt{2}\,\l\pp\,+\,m\r\,\pr \\ \\
\pl^2 &{~~}&\sqrt{2}\,\l\pp\,+\,m\r\,\pl &{~~}& \l\pp\,+\,m\r^2
\end{array}
\right]\,\,.\label{ob}
\eeq


For spin three half,  the matrix connecting the instant form hadronic wave
functions with the front form wave functions is
\def\tc{ {1\over {\left[2\,\l E\,+\,m\r\,\pp\,\right]^{3/2} }}  }
\beq
\Omega\l 3/2 \r \eq \tc \,
\left[
\begin{array}{cccc}
\beta(3/2)&{}&{}&0 \\ \\
0&{}&{}&\beta(3/2)
\end{array}
\right]\,,\label{ta}
\eeq
where the the $4\times4$ block matrix $\beta(3/2)$ is defined as
{
\FL
\begin{eqnarray}
&&\beta\l 3/2\r\eq \\ \nonumber
&&\left[
\begin{array}{ccccccc}
\l\pp\,+\,m\r^3 &{~}& -\,\sqrt{3}\,\l\pp\,+\,m\r^2\,\pr &{~}&
\sqrt{3}\,\l\pp\,+\,m\r\,\pr^2 &{~}& -\,\pr^3 \\ \\
\,\sqrt{3}\,\l\pp\,+\,m\r^2\,\pl &{~}& \left[\l\pp\,+\,m\r^2\,-\,
2\,\pr\,\pl\right]\,\l\pp\,+\,m\r &{~}&
-\left[2\,\l\pp\,+\,m\r^2-\pr\,\pl\right]\,\pr &{~}&
\sqrt{3}\,\l\pp\,+\,m\r\,\pr^2  \\ \\
\sqrt{3}\,\l\pp\,+\,m\r\,\pl^2  &{~}&
\left[2\,\l\pp\,+\,m\r^2-\pr\,\pl\right]\,\pl &{~}&
\left[\l\pp\,+\,m\r^2\,-\,
2\,\pr\,\pl\right]\,\l\pp\,+\,m\r &{~}&
-\,\sqrt{3}\,\l\pp\,+\,m\r^2\,\pr \\ \\
\pl^3 &{~}& \sqrt{3}\,\l\pp\,+\,m\r\,\pl^2  &{~}&
\sqrt{3}\,\l\pp\,+\,m\r^2\,\pl&{~}& \l\pp\,+\,m\r^3
\end{array}
\right]\,.\label{tb}
\end{eqnarray}
}

\def\c{{1\over {\left[2\,\l E\,+\,m\r\,p^+\,\right]^2}} }
\def\aa{\left(p^+\,+\,m\right)^4}
\def\ba{2\,\left({\pp}\,+\,m\right)^3\,\pl}
\def\ca{\sqrt{6}\,\left({\pp}\,+\,m\right)^2\,\pl^2}
\def\da{2\,\left({\pp}\,+\,m\right)\,\pl^3}
\def\ea{\pl^4}
\def\ab{-\,2\,\l{\pp}\,+\,m\r^3\,\pr}
\def\bb{2\left[\l E\,+\,m\r\,{\pp}\,-\,2\,\pr\,\pl\right]\,\l{\pp}\,+\,m\r^2}
\def\cb{\sqrt{6}\,\left[\l
E\,+\,m\r\,{\pp}\,-\,\pr\,\pl\right]\,\l{\pp}\,+\,m\r\,\pl}
\def\db{\left[6\,{\pp}\l E\,+\,m\r\,-\,4\,\pr\,\pl\right]\,\pl^2}
\def\eb{2\,\l{\pp}\,+\,m\r\,\pl^3}
\def\ac{\sqrt{6}\l{\pp}\,+\,m\r^2\,\pr^2}
\def\bc{\sqrt{6}\,\left[-\,\l E\,+\,m\r\,{\pp}\,+\,\pr\,\pl\right]
\l{\pp}\,+\,m\r\,\pr}
\def\cc{2\left[2\,{\pp}^2\,\l
E\,+\,m\r^2\,-\,3\,\l{\pp}\,+\,m\r^2\,\pr\,\pl\right]}
\def\dc{\sqrt{6}\left[\l
E\,+\,m\r\,{\pp}\,-\,\pr\,\pl\right]\,\l{\pp}\,+\,m\r\,
\pl}
\def\ec{\sqrt{6}\,\left({\pp}\,+\,m\right)^2\,\pl^2}
\def\ad{-\,2\,\l{\pp}\,+\,m\r\,\pr^3}
\def\bd{\left[6\,{\pp}\,\l E\,+\,m\r\,-\,4\,\pr\,\pl\right]\,\pr^2}
\def\cd{\sqrt{6}\,\left[-\,\l E\,+\,
m\r\,{\pp}\,+\,\pr\,\pl\right]\,\l{\pp}\,+ \,m\r\,\pr}
\def\dd{2\left[\l E\,+\,m\r\,{\pp}\,-\,2\,\pr\,\pl\right]\,\l{\pp}\,+\,m\r^2}
\def\ed{2\,\l{\pp}\,+\,m\r^3\,\pl}
\def\ae{\pr^4}
\def\be{-\,2\l{\pp}\,+\,m\r\,\pr^3}
\def\ce{\sqrt{6}\,\l{\pp}\,+\,m\r^2\,\pr^2}
\def\de{-\,2\,\l{\pp}\,+\,m\r^3\pr}
\def\ee{\l{\pp}\,+\,m\r^4}

For spin two, the matrix connecting the instant form hadronic wave functions
with the front form wave functions is
\beq
\Omega\l 2 \r \eq \c \,
\left[
\begin{array}{ccc}
\beta(2)&{}&0 \\ \\
0&{}&\beta(2)
\end{array}
\right]\,,\label{twoa}
\eeq
where the the $5\times5$ block matrix $\beta(2)$ is defined via the following
five
columns
\begin{eqnarray}
&&\beta(2)_{\alpha,1}\eq
\left[
\begin{array}{c}
\aa \\ \\
\ba \\ \\
\ca \\ \\
\da \\ \\
\ea
\end{array}
\right]\,,~~
\beta(2)_{\alpha,2}\eq
\left[
\begin{array}{c}
\ab \\ \\
\bb \\ \\
\cb \\ \\
\db \\ \\
\eb
\end{array}
\right]\,, \nonumber\\\nonumber \\
&& \beta(2)_{\alpha,3}\eq
\left[
\begin{array}{c}
\ac \\ \\
\bc \\ \\
\cc \\ \\
\dc \\ \\
\ec
\end{array}
\right]\,,\nonumber \\\nonumber \\
&& \beta(2)_{\alpha,4}\eq
\left[
\begin{array}{c}
\ad \\ \\
\bd \\ \\
\cd \\ \\
\dd \\ \\
\ed
\end{array}
\right]\,,~~
\beta(2)_{\alpha,5}\eq
\left[
\begin{array}{c}
\ae \\ \\
\be \\ \\
\ce \\ \\
\de \\ \\
\ee
\end{array}
\right]\,.\label{twob}
\end{eqnarray}

\newpage

\mediumtext
\begin{table}
\caption{Algebra associated with the stability group of the  $x^+\eq 0$ plane.
The Commutator  $\big[\,$Element in the {\it first} column,
Element in the {\it first} row$\,\big]\eq$The element at the
intersetction of {\it the} row and column.}
\begin{tabular}{clllllll}
\multicolumn{1}{c}{ } &\multicolumn{1}{c}{$P_1$}
&\multicolumn{1}{c}{$P_2$} &\multicolumn{1}{c}{$J_3$}
&\multicolumn{1}{c}{$K_3$} &\multicolumn{1}{c}{$P_{-}$}
&\multicolumn{1}{c}{$G_1$} &\multicolumn{1}{c}{$G_2$}\\
\tableline
 $P_1$ & $0$ & $0$ & $-iP_2$ & $0$ & $0$ & $iP_-$ & $0$\\
 $P_2$ & $0$ & $0$ & $iP_1$ & $0$ & $0$ & $0$ & $iP_-$\\
 $J_3$ & $iP_2$ & $-iP_1$ & $0$ & $0$ & $0$ & $iG_2$ & $-iG_1$\\
 $K_3$ & $0$ & $0$ & $0$ & $0$ & $iP_-$ & $iG_1$ & $iG_2$\\
 $P_-$ & $0$ & $0$ & $0$ & $-iP_-$ & $0$ & $0$ & $0$\\
 $G_1$ & $-iP_-$ & $0$ & $-iG_2$ & $-iG_1$ & $0$ & $0$ & $0$\\
 $G_2$ & $0$ & $-iP_-$ & $iG_1$ & $-iG_2$ & $0$ & $0$ & $0$\\
\end{tabular}
\end{table}

\end{document}